\documentstyle[preprint,aps,eqsecnum]{revtex}
\begin{document}

\title {Weak localization effect on thermomagnetic phenomena}
\author {Michael Reizer}
\address{5614 Naiche Rd. Columbus, Ohio 43213}
\author {Andrew Sergeev}
\address{ Institut  f\"ur Theoretische  Physik,  Universit\"at  Regensburg, 
D-93040 Regensburg,  Germany }
\maketitle

\begin{abstract}
\noindent

The quantum transport equation (QTE) is extended to study
weak localization (WL) effects on galvanomagnetic and thermomagnetic phenomena.
QTE has many advantages over the linear response method (LRM):
(i) particle-hole
asymmetry which is necessary for the Hall effect is taken into account
by the nonequilibrium distribution function, while LRM requires
expansion near the Fermi surface, (ii) when calculating response to
the temperature gradient, the problem of WL correction to the 
heat current operator is avoided, 
(iii) magnetic field is directly introduced to 
QTE, while the LRM deals with the vector potential
and and special attention should be paid to maintain gauge invariance,
e.g. when calculating the Nernst effect the heat current operator
should be modified to include the external magnetic field. 
We reproduce in a very compact form known results for 
the conductivity, the Hall and the thermoelectric effects and 
then we study our main problem,
WL correction to the Nernst coefficient (transverse thermopower).
We show that in a quasi-two-dimensional film
the Nernst coefficient has a large
logarithmic factor similar to that of the conductivity and the
Hall conductivity, while the thermoelectric coefficient does not have
such a factor.

\end{abstract}
\pagebreak
 
\centerline{\bf 1. Introduction} 
In many complex cases the
quantum transport equation (QTE) turns
out to be physically clear and more convenient than the linear response
method (LRM). 
Calculating many-body corrections to 
the electrical conductivity choice of QTE or LRM is
a matter of taste, LRM requires many diagrams to be considered,
while QTE deals only with the electron self-energy diagrams 
but includes specific terms like Poisson brackets corrections.
QTE shows special advantages when thermoelectric
and galvanomagnetic effects are considered:
 
1. Calculating the thermoelectric coefficient by QTE as a
response of the electron system to the temperature gradient, 
one avoids difficulty associated  with corrections to the 
heat current operator due
to the electron-electron and electron-phonon interactions \cite{RS,RSWL}. 

2. Calculating the Hall coefficient by QTE the
electron-hole asymmetry is accounted automatically by the nonequilibrium
distribution function of noninteracting electrons, while LRM
requires all electron parameters to be expanded near the Fermi surface,
 
3. QTE incorporates real electric and magnetic fields in the gradient 
terms and also through the terms in form of the Poisson brackets.
Contrary to that LRM deals with two vector potentials for the 
electric and magnetic fields and special attention requires 
to obtain the Hall component of the electric current. Besides that 
when calculating the Nernst coefficient the gauge invariance requires the 
vector potential corresponding to the magnetic field to be included
in the heat current operator (we will discuss this subject in detail
in Appendix A)

QTE was initially developed for the interaction effects on the conductivity
in the diffusion regime \cite {A} and was later extended to
the weakly disordered regime for the conductivity \cite{RS1},
thermoelectric power \cite{RS,RSWL}. and the Hall effect \cite{R}.
 In the present paper we first derive
QTE which incorporates electric and magnetic fields and the
temperature gradient on equal footing. Then we obtain the WL
corrections to the conductivity, the Hall conductivity
which have been calculated earlier by LRM in 
\cite {AAL} and \cite{F} ( see also \cite {AKLL} and \cite{AA}).
Then we consider the thermoelectric
coefficient which have also been considered earlier by LRM \cite{AGG}.
Finally we find WL correction to the Nernst coefficient, 
the problem which have not been studied before, and for the reasons
mentioned above, it is a very difficult problem to study by LRM.

\centerline{\bf 2. Quantum transport equation}
The quantum transport equation method based on the Keldysh
diagrammatic technique, 
where transport phenomena are described by $2\times 2$
matrix electron Green function ${\hat G}$, as well as
the matrix electron self energy $ {\hat \Sigma}$,
$$ {\hat G} =\pmatrix{0 & G^A \cr G^R & G^C \cr}, \ \ \ \ \ \ \ \
{\hat\Sigma}=\pmatrix{\Sigma^C & \Sigma^R \cr \Sigma^A & 0}, \eqno(1)$$
where $A$ and $R$ stand for advanced and retarded components of the matrix
function and $C$ corresponds to the kinetic component.

The electric current is expressed by the kinetic Green's function $G^C$,
$${\bf J}=
=\int{d^{4}P\over (2\pi)^{4}}e{\bf v}{\rm Im}\Delta G^C(P),     \eqno(2)$$
where $P=({\bf p},\epsilon)$, and $\Delta G^C$ is the 
nonequilibrium correction to $G^C$, which will be calculated in
each particular case of external disturbance.

Assuming that weak localization corrections are small, 
we will calculate the kinetic electron Green function by iteration.
Without weak localization effects, the retarded (advanced) 
Green function is
$$G^R(P)=[G^A(P)]^*={1\over \epsilon-\xi_p+i/2\tau},\ \ \ \xi_p={p^2-p^2_F
\over 2m},                                                       \eqno(3)$$
where $\tau$ is the elastic scattering time due to 
electron-impurity scattering.

The normalization condition for the matrix Green function 
in the coordinate representation
$$  \int dY {\hat G}(X_2,Y){\hat G}(Y,X_1)={\hat 1} ,            \eqno (4)$$
($X$ is a four dimensional coordinate $({\bf r},t)$),
results in following form of the Keldysh component
$$  G^C(X_2,X_1)=\int dY \ [S(X_2,Y)G^A(Y,X_1)-G^R(X_2,Y)S(Y,X_1)], \eqno(5)$$
where the function S plays the role of the electron
distribution function.

In equilibrium $S=S_0=-\tanh(\epsilon/T)$. In the presence of the electric
and weak magnetic fields (the quantization of the electron levels 
is neglected) $S$ is 
determined from the following transport equation:
$$e({\bf v}\cdot{\bf E} ){\partial S\over \partial \epsilon}
-({\bf v}\cdot{\nabla  T} ){\epsilon\over T}
{\partial S\over \partial \epsilon}
+{e\over c}({\bf v}\times{\bf H}){\partial S\over \partial {\bf p}}
=I_{e-imp},                                                         \eqno(6)$$
where $I_{e-imp}$ is the collision integrals corresponding
to the electron-impurity interaction in the lowest order 
(without WL corrections) is chosen in the simplest form:
$$I_{e-imp}={2\over \pi \nu \tau} \int {d{\bf k} \over (2\pi)^3}[S({\bf k},
\epsilon)-S({\bf p},\epsilon)]{\rm Im}G^A_0({\bf k},\epsilon)=
{S_0(\epsilon)-S(\epsilon)\over \tau}.                             \eqno(7)$$

Performing the Fourier transformation of Eq. (5) from the coordinate
representation to the momentum-energy representation we get
$$  G^C(P)= S(P)[G^A(P) 
-G^R(P)] \ + \ \delta G^C(P),                                  \eqno (8)$$
where $\delta G^C$ 
correction in the form of the Poisson bracket
$$ \delta G^C = {i\over 2}\lbrace S(P),
G^A(P)+G^R(P)\rbrace.                                           \eqno(9)$$

Poisson bracket corrections arise when integrals in the coordinate
representation (like Eq. (5)) transform to the momentum-energy representation,
$$ \int dX \ A(X_1,X)B(X,X_2) \Rightarrow  
 A(P)B(P)
+{i \over 2}\lbrace  A(P),B(P) \rbrace,                            \eqno(10)$$
where 
$$ \lbrace A,B \rbrace =\biggl( {\partial 'A \over \partial \epsilon }
                    {\partial 'B \over \partial t}-
                   {\partial 'B \over \partial \epsilon }
                  {\partial 'A \over \partial t }\biggr)
                - \biggr( {\partial 'A \over \partial {\bf p} }
                {\partial 'B \over \partial {\bf r} }
                - {\partial 'B \over \partial {\bf p} }
               {\partial 'A \over \partial {\bf r} }  \biggl).  \eqno(11)$$
The potentials ${\bf A}$ and $\Phi$ enter as 
$$ {\partial ' \over \partial t}= {\partial \over \partial t}
   -{e\over c}{\partial {\bf A} \over \partial t}
{\partial \over \partial {\bf p} }
-e{\partial \Phi \over \partial t} {\partial \over \partial \epsilon}, $$
$$ {\partial ' \over \partial {\bf r} }={\partial \over \partial {\bf r}} -
{e\over c}{\partial {\bf A} \over \partial r_i } {\partial \over \partial 
p_i } -e{\partial \Phi \over \partial {\bf r} }{\partial  \over \partial
\epsilon }.                                                      \eqno(12)$$
Therefore Poisson brackets due to the electric and magnetic fields, and 
also due to the temperature gradient are 
$$\lbrace A,B\rbrace_E =e{\bf E}\biggl({\partial A\over \partial \epsilon}
{\partial B\over \partial{\bf p}}-
{\partial B\over \partial \epsilon}{\partial A\over \partial{\bf p}}\biggr), 
                                                                 \eqno(13)$$
$$\lbrace A,B\rbrace_H={e\over c}{\bf H}\cdot
\biggl({\partial A\over \partial {\bf p}}\times{\partial B\over \partial
{\bf p}}\biggr),                                                 \eqno(14)$$
$$\lbrace A,B\rbrace_T =\nabla T\biggl({\partial A\over \partial T}
{\partial B\over \partial{\bf p}}-
{\partial B\over \partial T}{\partial A\over \partial{\bf p}}\biggr), 
                                                                 \eqno(15)$$

Our goal is to find the WL correction $\Delta G^C$ and  calculate 
the electric current (Eq. (6)). The kinetic Green function, $G^C$
may be found by two methods. First, $G^C$ may be expressed through 
the nonequilibrium electron distribution function 
(see Eq. (5)), which in turn is determined from 
the transport equation \cite{RS,RSWL,A}. Second, the transport equation 
may be written directly for $G^C$. 
Both methods are equivalent, we chose the second method as
more convenient for the current problem.

Note, that retarded (advanced) component of the electron self-energy 
consists of integrals of retarded (advanced) Green functions,
therefore, in the main approximation it is equal to zero. 
Thus, WL correction to $G^C$ is determined only
by the  kinetic component  $\Sigma^C$. In the coordinate
representation the transport equation for $G^C$ is
$$ \Delta G^C (X,X') 
= \int \ dY \ dZ \ G^R(X,Y) \Sigma^C(Y,Z) G^A(Z,X').              \eqno(16)$$
For the problem of WL the self-energy diagrams $\Sigma$ are shown in Fig. 1.
To solve Eq. (16) we use iteration procedure. 
As in the case of QTE for the nonequilibrium electron distribution function
\cite{RS,RSWL,A},
field terms and scattering terms in Eq. (16) cannot be separated.
In the present formalism, field terms
accounting for the electric (magnetic) field 
and temperature gradient appear due to zero order (without WL corrections) nonequilibrium distribution functions obtained from
Eq. (6) and due to Poisson brackets (Eqs. (13-15)) after transformation of 
Eq. (16) to the energy-momentum representation.
Further for  each particular external perturbation 
the procedure described above will be presented in detail. 

\centerline{\bf 2. Conductivity}

For the conductivity we retain in Eq. (6) only the term corresponding
the electric field.  
Solution of Eq. (6) with the electron-impurity collision integral, Eq. (7),
in the energy-momentum presentation it is given by
$$S=S_0+\phi_E, \ \ \                          
\phi_E(P)=-e\tau ({\bf v}\cdot {\bf E})
{\partial S_0(\epsilon)\over \partial \epsilon}.                \eqno(17)$$

The Poisson bracket term for 
$G^C$ (Eq. (9)) is 
$$ \delta_E G^C = {i\over 2}\lbrace S_0(\epsilon),
G^A(P)+G^R(P)\rbrace_E.                                           \eqno(18)$$

Now we take into account WL corrections for $G^C$, using Eq. (16).
This equation in the momentum representation 
with all terms in the form of the Poisson brackets is

$$ \Delta G^C  =  G^R \Sigma^C(\phi_E) G^A 
+ G^R \Sigma^C(\delta_E G^C) G^A 
 + G^R \delta_E\Sigma^C(S_0) G^A $$
$$+ {i\over 2} \{G^R,\Sigma^C(S_0) \}_E  G^A 
+ {i\over 2} G^R \{\Sigma^C(S_0), G^A \}_E. 
                                                                  \eqno(19)$$
First three terms take into account the nonequilibrium corrections
due to $\phi_E$, $\delta G^C$, and internal Poisson brackets in
the electron self-energy $\Sigma^C$. Last two terms are the Poisson
brackets between the self-energy and external Green functions.   
Keeping the linear in the electric field terms,
one should calculate the Poisson bracket terms  
with the equilibrium electron self-energy. But 
as we have already mentioned, $\Sigma^A =0$, and therefore
$\Sigma^C(S_0)=2iS_0(\epsilon){\rm Im}\Sigma^A=0 $. 
So the last two terms are equal to zero, 
and we need to take into account only  
nonequilibrium corrections to the electron self-energy.
  
Let us consider  diagrams shown on Fig. 1. We denote corresponding
self-energies as $\Sigma_1$, $\Sigma_2$, and  $\Sigma_3$.
The nonequilibrium corrections to the first diagram are
$$\Sigma^C_1 (\phi_E) =\phi_E(Q-P)C(Q)[G^A(Q-P)-G^R(Q-P)] , $$
$$\Sigma^C_1(\delta_EG^C)= C(Q)\delta_E G^C(Q-P) ,                \eqno(20)$$
where for brevity we dropped the integral 
$\int dQ/(2\pi)^4$,$Q=({\bf q},\omega)$, The Cooperon, $C(Q)$, is
$$C(Q)={1\over \pi\nu\tau^2}{1\over -i\omega+ 1/\tau_\phi+ Dq^2},  \eqno(21)$$
where $\tau_\phi$ is the phase relaxation time.

The first correction in Eq. (19) $\Delta_1 G^C$ due to 
$\Sigma_1(\phi_E)$ is given by
$$  \Delta_1G^C (P)=G^R(P)C(Q)\phi_E(P)[G^A(Q-P)-G^R(Q-P)]G^A(P). \eqno(22)$$
Substituting  $\Delta_1G^C (P)$ into Eq. (2)
we find corresponding correction to the conductivity. 
In the quasi-two-dimensional case, $d<<L_\phi$, where 
$d$ is the thickness of the film and $L_\phi =(D\tau_\phi)^{1/2}$ 
is the localization length, we have
$$\Delta_1\sigma_{xx}=-{2e^2D\over \pi d}\int{d^2q\over (2\pi)^2}{1\over
1/\tau_\phi+Dq^2}=-{e^2\over 2\pi^2d}
\rm ln \bigg( {L_\phi \over \ell} \bigg),                           \eqno(23)$$
where $\ell=v_F\tau$ is the mean free path.

The second correction in Eq. (19) $\Delta_2G^C$ due to $\Sigma_1(\delta G^C)$
(Eq. 20) is 
$$  \Delta_2G^C (P) = G^R(P) C(Q) (i/2) \ \{S_0(\epsilon),
 G^A(Q-P)+G^R(Q-P)\}_E G^A(P),                              \eqno(24)$$
which results in the correction to the conductivity
$\Delta_2\sigma_{xx}=(1/2)\Delta_1\sigma_{xx}$.

Now we consider diagrams  $\Sigma_2$ and $\Sigma_3$. 
These diagrams contain 
$G^C$ inside the electron-impurity ladder. 
For this reason, nonequilibrium corrections due to the distribution function 
$\phi_E$ and due to the Poisson bracket $\delta_E G^C$ 
give zero after the angular integration over the electron momentum.

The next correction, $\Delta_3G^C$, comes from 
the Poisson brackets inside the self-energy diagrams.
The Poisson bracket should include the electron distribution 
function $S_0$ in $G^C$.
As total contribution of all three equilibrium diagrams is zero,
and in all diagrams $G^C$ has the same position with respect 
to the Cooperon,
the Poisson brackets between $S_0(\epsilon)$ and 
Green's functions inside the Cooperon are equal to zero.
The only nonzero term in $\Delta_3G^C$ appears when 
the Poisson brackets are taken between $S_0(\epsilon)$ and 
the electron Green functions $G^{A,R}(Q-P)$  
(external with respect to the Cooperon) in the second and third diagrams,
$$ \Delta_3G^C= G^R(P)\biggl[-{i\over2}C(Q)\lbrace S_0(\epsilon),
G^A(Q-P)+G^R(Q-P)\rbrace_E\biggr]G^A(P),                          \eqno(25)$$
which results in $\Delta_3\sigma_{xx}= -(1/2)\Delta_1\sigma_{xx}$.
Thus, the second and third corrections to the conductivity
originating from the Poisson brackets mutually cancel out and 
$\Delta_1\sigma_{xx}$ describes the full WL correction to the conductivity
which coincides with \cite{AA}.
However, we will show that the Poisson brackets corrections
are very important for the Hall and Nernst effects.

\centerline{\bf 3. Hall effect}

Calculating the Hall conductivity 
we assume that the magnetic field is directed along the $z$-axis and
the electric field is directed along the $x$-axis. The Hall current $J_y$ is
proportional to ${\bf E}\times{\bf H}$. 

Without WL effects the electron distribution function has the form
$$ S=S_0+\phi_E+\phi_H,                                    \eqno(26)$$
where the nonequilibrium distribution function $\phi^E$ is given
by Eq. 17, and $\phi_H$ is
$$\phi_H({\bf p},\epsilon)=
-\tau{e\over c}({\bf v}\times{\bf H})
{\partial (S_0+\phi_E)\over \partial {\bf p}}
=-{e^2\tau^2\over cm}{\bf v}\cdot({\bf E}\times{\bf H})
{\partial S_0\over \partial \epsilon}.                           \eqno(27)$$
The Hall conductivity of noninteracting electrons is given by 
$$\sigma_{xy}={J_y\over E}=2e\int{d^{4}P\over (2\pi)^{4}}v_y\phi_H
{\rm Im}G^A(P)=\Omega\tau\sigma_{xx},                             \eqno(28)$$
where $\Omega=eH/mc$ is the cyclotron frequency and $\sigma_{xx}$ is
the Drude conductivity.

It is noteworthy here, that the Hall effect is proportional to
the particle-hole asymmetry which is taken into account in the 
nonequilibrium distribution function in Eq. (27)
by the factor $\partial^2 \xi_p / \partial^2 p = 1 /m $, which is a measure
of the average curvature of the Fermi surface.

For electrons in the magnetic field the Poisson bracket corrections 
to $G^C$, 
$$ \delta_H G^C = {i\over 2}\lbrace \phi_E(P),
G^A(P)+G^R(P)\rbrace_H.                                           \eqno(29)$$
is also proportional
to the same particle-hole asymmetry factor as $\phi_H({\bf p},\epsilon)$. 

To get WL correction to the  Hall
conductivity we keep in the momentum representation of Eq. (16) 
only terms proportional to ${\bf E}\times{\bf H} $, 
$$ \Delta G^C  =  G^R\Sigma^C(\phi_H) G^A+ G^R\Sigma^C(\delta_HG^C) G^A
+G^R \delta_H \Sigma^C(\phi_E)G^A
+ G^R (\delta_H \delta_E \Sigma^C(S_0)) G^A $$                     
$$+ {i\over 2} \{G^R ,\Sigma^C(\phi_E) \}_H  G^A 
+ {i\over 2} G^R  \{\Sigma^C(\phi_E), G^A \}_H,                \eqno(30) $$
where $\Sigma^C(\phi_H)$ stands for the self-energy with nonequilibrium 
distribution function $\phi_H$,
$\Sigma^C(\delta_HG^C)$ stands for the self-energy with the magnetic Poisson
bracket correction to $G^C$ from Eq. (29),
$\delta_H \Sigma^C(\phi_E)$ denotes the magnetic Poisson bracket in
the self-energy with the nonequilibrium function $\phi^E$,
$\delta_H \delta_E \Sigma^C$ is the correction due to
internal (inside  $\Sigma^C$), double Poisson brackets. 
In the last two terms we calculate
the magnetic Poisson bracket between the nonequilibrium self-energy
and $G^{R,A}$ functions. 
As we discussed in the previous section, in equilibrium $\Sigma^C=0$,
thus external electric Poisson brackets between  $\Sigma^C(S_0)$
and $G^R$ and $G^A$ are equal to zero. 

Let us  start with the first diagram,  $\Sigma_1$.
The correction $\Delta_1G^C$ corresponding to the first term in Eq. (30)
is
$$\Delta_1G^C(P) = G^R(P)\Sigma^C(\phi_H)G^A(P)$$
$$=G^R(P) C(Q) \phi_H(P)[G^A(Q-P)-G^R(Q-P)]G^R(P).                  \eqno(31)$$
Calculating the electric current, we find the correction 
to the Hall conductivity $\Delta_1\sigma_{xy}=\Omega\tau\delta \sigma_{xx}$.

From the second term in Eq. (30 ) we have
$$\Delta_2G^C(P) = G^R(P)\Sigma^C(\delta_HG^C(Q-P))G^A(P)$$
$$=-{i\over 2 \tau } G^R(P)\phi_H (P)
 [(G^A(Q-P))^2+(G^R(Q-P))^2]G^A(P).                             \eqno(32)$$
The corresponding correction to the Hall conductivity is
$\Delta_2\sigma_{xy}=(1/2)\Omega\tau\delta \sigma_{xx}$.

Note that, as for the conductivity, the corrections to the Hall conductivity
from $\phi_H$ and $\delta_HG^C$ in $\Sigma_2$ and $\Sigma_3$ 
give zero after the angular integration.

Now we calculate the internal Poisson brackets (third term in Eq. (30)).
First let us
consider the magnetic Poisson brackets between $\phi_E$ and 
and external
with respect to the Cooperon Green functions  $G^{A,R}(Q-P)$ 
in the second and the third diagrams.
For the Hall
conductivity these contributions
cancel each other, because the electron momentum in 
$\phi_E$ changes its sign in these diagrams.
Terms with the magnetic Poisson bracket  between $\phi_E$ and the 
Cooperon for all three diagrams $\Sigma_1$,  $\Sigma_2$, and  
$\Sigma_3$ are proportional to
$$ [G^A(Q-P)-G^R(Q-P)]+ G^A(Q-P)<G^A(P')-G^R(P'), G^A(Q-P')>  $$
$$+ G^R(Q-P)<G^R(P'),G^A(Q-P')-G^R(Q-P')>,                      \eqno(33) $$
where $<G_1, G_2>$ stands for
$(\pi \nu \tau)^{-1} \int d {\bf p} G_1 G_2$.
Performing calculations, we find that these terms cancel each other.

Similar way one can prove that the double Poisson brackets corrections 
$\delta_H\delta_E \Sigma^C$ also mutually cancel out.

Now we calculate contributions of the fifth and sixth terms in Eq. (30). 
$$ \Delta_5G^C+\Delta_6G^C = {e\over c} {\bf H}\cdot {\bf v} \times 
{\partial \Sigma^C \over \partial {\bf p}} [(G^A)^2G^R-G^A(G^R)^2], \eqno(34)$$
Calculations show that only $\Sigma_2$ and  $\Sigma_3$ give
nonzero contribution, thus
$$ \Delta_5G^C+\Delta_6G^C = {e\over c} {\bf H}\cdot {\bf v} \times 
{\partial \over \partial {\bf p}} [\Sigma^C_2(\phi_E) + 
\Sigma^C_3(\phi_E)] [(G^A)^2G^R-G^A(G^R)^2]$$
$$={e\over c} {\bf H}\cdot {\bf v} \times {\bf E}
C(Q) {e \tau \over m} 
{\partial S_0\over \partial \epsilon}
[G^A(Q-P)+G^R(Q-P)][(G^A)^2G^R-G^A(G^R)^2].                \eqno(35)$$
Corresponding correction to the Hall conductivity is
$\Delta_{5,6}\sigma_{xy}=\sigma_{xy}(1/2) \Omega\tau \delta \sigma_{xx}$.

Finally, the total contribution to the Hall conductivity is  
$${\Delta\sigma_{xy}\over \sigma_{xy}}=
2{\Delta\sigma_{xx}\over \sigma_{xx}},                         \eqno(36)$$
which coincides with \cite{AAL} and \cite{F}.

\centerline{\bf 4. Thermoelectric and Nernst effects}

Now we consider the thermoelectric and Nernst coefficients.
Without WL correction, the nonequilibrium distribution 
function in the crossed temperature gradient and the magnetic field is
$$S=S_0+\phi_T+\phi_N,                         \eqno(37)$$   
$$\phi_T(P)=\tau(\epsilon) ({\bf v}\cdot {\nabla T}){\epsilon\over T}
{\partial S_0(\epsilon)\over \partial \epsilon},                 \eqno(38)$$
$$\phi_N({\bf p},\epsilon)=
-\tau{e\over c}({\bf v}\times{\bf H})
{\partial (S_0+\phi_T)\over \partial{\bf p}}
={e\tau^2(\epsilon)\over cm}{\bf v}\cdot({\nabla T}\times{\bf H})
{\epsilon\over T}{\partial S_0\over \partial \epsilon},         \eqno(39)$$
The Poisson-bracket corrections to the $G^c$ are
$$ \delta_T G^C = {i\over 2}\lbrace S_0(\epsilon),
G^A(P)+G^R(P)\rbrace_T,                                           \eqno(40)$$
$$ \delta_N G^C = {i\over 2}\lbrace \phi_T(P),
G^A(P)+G^R(P)\rbrace_H.                                           \eqno(41)$$

The thermoelectric coefficient is defined from the equation
$J=\eta\nabla T$, therefore substituting $\phi_T$ from Eq. (38) to
Eq. (2) we have the following equation for the thermoelectric
coefficient without weak localization effect
$$\eta_0={2e\over T}\int{d^4P\over (2\pi)^4}
 {\bf v}^2 \tau(\epsilon)\epsilon {\partial S_0(\epsilon)\over 
\partial \epsilon}{\rm Im}G^A(P).                             \eqno(42)$$
To get a nonzero result from Eq. (42) requires the electron-hole
asymmetry to be taken into account. Therefore we expand all electron
quantities near the Fermi surface, 
$$v(\epsilon)=v_F\biggl(1+{\epsilon\over 2\epsilon_F}\biggr),\ \  
\nu(\epsilon)=\nu_0\biggl(1+{\epsilon\over 2\epsilon_F}\biggr),\ \ 
\tau(\epsilon)=\tau_0\biggl(1-{\epsilon\over 2\epsilon_F}\biggr),\ \ \nu_0={mp_F\over \pi^2},                                   \eqno(43)$$
where $n_i$ is impurity concentration and 
$U$ is the electron-impurity potential.
Finally we have
$$\eta_0=-{2\over 9}e\tau_0p_FT.                           \eqno(44)$$

The weak localization correction to the thermoelectric
coefficient is obtained by substituting $\phi_T$ for $\phi_E$ and
$ \lbrace A,B \rbrace_T$ for  $\lbrace A,B \rbrace_E$ to 
Eqs. of Sec. 3. As a result we get
$$\Delta\eta=-{e\over \pi T}\int{d^3q\over (2\pi)^3}
\int d\epsilon\epsilon {\partial S_0(\epsilon)\over \partial \epsilon}
{1\over q^2 + L_\phi^{-2}(\epsilon)},                            \eqno(45)$$
where $L_\phi^2(\epsilon) = D(\epsilon)\tau_\phi(\epsilon)$.
For a film, which is two-dimensional with respect to WL ($d<< L_\phi$),
the momentum integral in Eq. (45) is logarithmically divergent on
the upper limit. Taking the cutoff to be $1/\ell(\epsilon)$, 
$\ell(\epsilon)=v(\epsilon)\tau(\epsilon)$,
and making the expansion near the Fermi surface, we get
$$\Delta\eta=-{e\over 2 \pi^2 Td} 
\int d\epsilon  \ \epsilon {\partial S_0(\epsilon)\over \partial \epsilon}
{\rm ln} \biggl( {L_\phi(\epsilon) \over \ell(\epsilon)} \biggr)
={e^2T\over 12d}\biggl({\partial\over \partial\epsilon}{\rm ln} 
\bigg({L_\phi(\epsilon)\over\ell(\epsilon)}\bigg)\biggl)_{\epsilon=0},
                                                            \eqno(46)$$
which coincides with \cite{AGG}. 
Equation (46) shows that the correction to the thermoelectric coefficient 
does not have a large logarithmic factor compare to the
correction to the conductivity, Eq. (23).
This result means that for experimentally measured Seebek coefficient
$$S=S_0\biggl(1-{\Delta\sigma_{xx}\over \sigma_{xx}}
+{\Delta\eta\over \eta_0}\biggr),                        \eqno(47)$$
where $\eta_0=-\eta_0/\sigma_{xx}$, only the term corresponding
to the correction to the conductivity is important.

The Nernst coefficient, $N$, is defined by the equation 
${\bf J}=N(\nabla T \times {\bf H})$.
Substituting $\phi_N$ for $\phi_H$ in Eq. (29) we get for the Nernst
coefficient without weak localization effects
$$N=-{\pi^2\over 9}{e^2T\over cm}{\partial \over \partial \epsilon}
\biggl(v^2(\epsilon)\tau^2(\epsilon)\nu(\epsilon)\biggr)|_{\epsilon=0}= 
-{\pi^2\over 6}{T\over \epsilon_F}(\Omega\tau_0){\sigma_{xx}\over H}=
-{\pi^2\over 6}{T\over \epsilon_F}\sigma_{xy}={\Omega \tau_0 \over 2H}
\eta.                                                            \eqno(48)$$   
Equation (48) shows that the Nernst coefficient 
is proportional to the product of the electron-hole
asymmetry factors, one from the Hall conductivity $\sigma_{xy}$ 
and another factor $T/\epsilon_F$ arises after expansion of all electron  
parameters near the Fermi surface similar to the 
thermoelectric coefficient. 
Note that this important fact is clear when calculating
the Nernst coefficient by the transport equation. When LRM is applied
special attention requires to find this double electron-hole
asymmetry, as discusses in detail in Appendix A.

The WL correction to the Nernst coefficient may be 
obtained from the presented above calculation of the Hall conductivity
by substituting  $\phi_T$ for $\phi_E$ and also
$\lbrace A,B \rbrace_T$ for $\lbrace A,B \rbrace_E$ in all equations
of Sec. 3. As a result we have
$$\Delta N=-{2e^2\over \pi cm}\int{d^3q\over (2\pi)^3}
\int d\epsilon \ {\epsilon \over T} 
{\partial S_0(\epsilon) \over \partial \epsilon} 
{\tau(\epsilon) \over q^2 + [D(\epsilon)\tau_\phi(\epsilon)]^{-1}}.\eqno(49) $$
According to Eq. (49), in the quasi-two-dimensional case ($d< L_\phi$) 
the main contribution to the Nernst coefficient arises from the 
expansion of $\tau(\epsilon)$ (see Eq. (43)), as a result
$$ {\Delta N \over N} = - 2 {\Delta \sigma_{xx} \over \sigma_{xx}}, \eqno(50)$$
which is different by sign compared with corresponding correction to 
the Hall conductivity (Eq. (36)).
Equation (50) shows that in the quasi-two-dimensional case 
the WL correction to the Nernst coefficient has a large logarithmic factor.
Note, that for exactly 2D electrons, the electron momentum
relaxation time does not depend on the energy, and the large 
logarithmic factor in WL correction to the Nernst effect disappears.

\centerline{\bf 5. Conclusions}

In the present paper we apply QTE for WL effect
on different transport coefficients. We
reproduce WL corrections to the conductivity, 
the Hall conductivity \cite{AAL}, {\cite{F}  and the 
thermoelectric effect \cite{AA} obtained earlier by LRM.
Then we calculate WL corrections to the Nernst coefficient, 
this problem have not been considered before.

QTE gives convenient and universal description
of all transport phenomena. Specific feature of an external
disturbance is taken into account by the nonequilibrium 
distribution functions of noninteracting 
electrons (Eqs. (17), (27), (38), and (39)) and by 
quantum corrections to the quasiclassical transport equation in
the form of the Poisson brackets (Eqs. (13), (14), and (15)). 

Quasiclassical transport equation has been applied to the WL effect on
the electrical conductivity in \cite{RmS}, where all terms
in the form of the Poisson brackets have been ignored. 
As we demonstrated above, these terms
mutually canceled for the correction to the conductivity but 
they are very important for the Hall conductivity and the Nernst coefficient.

QTE also allows us to avoid lengthy calculations of the particle-hole
asymmetry, which is important for the Hall and Nernst effects.
In the Hall effect the particle-hole asymmetry appears as 
a measure of the average curvature of the Fermi surface, and 
is expressed through the electron mass (some effective electron mass in 
the general case). Using QTE we get
the Hall's particle-hole asymmetry in the distribution function
of noninteracting electrons, Eq. (27), and in the magnetic Poisson bracket,
Eq. (14). To obtain the same result by LRM
one needs to expand all electron parameters in all Green's
functions near the Fermi surface.

The Nernst coefficient is proportional to the double particle-hole 
asymmetry. It means that in the linear response method one should 
keep second-order terms in expansion of electron parameters near
the Fermi surface. With many-body corrections, it is a complicated 
problem. Ignoring at least one of the diagram gives a nonzero
Nernst coefficient in zero order in the particle-hole asymmetry (without
any expansion).  We discuss this problem in detail in Appendix A
for noninteracting electrons. For interacting electrons the problem
is even more complicated. As an example of the problem for which
wrong results have been obtained, we mention the effect 
of superconducting fluctuations on the Nernst coefficient \cite{MAK,DOR},
(see \cite{AVS}for more details).

We also found that in the quasi-two-dimensional case  
WL correction to the Nernst
coefficient is proportional to a large logarithmic factor similar
to WL corrections to the conductivity and the Hall conductivity,
while for WL correction to the thermoelectric coefficient 
such a correction is absent. 

Thermoelectric phenomena in low-dimensional systems attracted considerable
attention recently, \cite{Flet,ZBK,SK,MFZ}. We hope that experimental
study of the Nernst coefficient may improve our understanding of
kinetic phenomena in low-dimensional and mesoscopic systems.

\vskip 1cm
\centerline{\bf Acknowledgments}
\vskip 0.3cm
The support by the Alexander von Humboldt
Foundation (AS) is greatly acknowledged.
\vskip 1cm

\centerline{\bf Appendix A}

In this appendix
we calculate the Nernst coefficient of noninteracting 
electrons using LRM. We demonstrate that
the heat current correction due to the magnetic field should be taken
into account. The diagram with this additional heat-current vertex 
cancels the basic diagrams in the zero order in the electron-hole
asymmetry. The nonzero Nernst coefficient
arises only in the second order in the asymmetry.
These conclusions are also relevant to many-body corrections
to the Nernst coefficient.

In the linear response method the external magnetic field
is introduced through the corresponding vector potential,
${\bf H}=i({\bf k} \times {\bf A})$. 
In the presence of the external magnetic field the electronic
kinetic energy is given by
$$ K = {1 \over 2m} \bigg( {\bf p} + {e \over c} {\bf A} \bigg)^2=
{p^2 \over 2m}+{e\over c} ({\bf v} \cdot {\bf A})
+{e^2\over 2mc^2}A^2.                                       \eqno(A.1)$$

We choose the gauge:
$ {\bf k} \cdot {\bf A} = 0$, and 
consider a response of the electron system to the temperature gradient, 
which is perpendicular to the magnetic field.

Then the Nernst current 
is proportional to 
$ \nabla T \times {\bf H} = -i{\bf A}({\bf k} \cdot \nabla T)$.

In LRM the Nernst coefficient is given by
$$N={1\over \Omega TH}{\rm Im}Q^R(\Omega,{\bf A}).               \eqno(A.2)$$
Here $Q^R(\Omega,{\bf A})$ is the Fourier representation of the 
retarded correlation function of the heat and charge currents
operators $\hat{\bf J}_h$ and $\hat{\bf J}_e$
in the presents of vector potential ${\bf A}$,
$$Q^R({\bf x}-{\bf x'},t-t',{\bf A})
=-\Theta(t-t')<[\hat{\bf J}_h({\bf x},t,{\bf A})\cdot{\bf n},
\hat{\bf J}_e({\bf x'},t',{\bf A})\cdot{\bf n_1}]>,              \eqno(A.3)$$
where ${\bf n}$ is a unit vector directed along the temperature
gradient,  ${\bf n}$ is a unit vector perpendicular to ${\bf n}$ and
the external magnetic field ${\bf H}$.

The external magnetic field is taken into account by inserting 
additional the magnetic vertex
$ (e/c) ({\bf v} \cdot {\bf A})$ (the second term on the right-hand
side of Eq. (A.1) into electronic Green's function of the corresponding
diagrams of the linear response correlators.  
For example, diagrams for the Hall coefficient 
are obtained  by inserting this vertex
into the conductivity diagrams as it was done in \cite{AKLL} and \cite{AA}. 

For the Nernst coefficient, however it is not enough to insert the
magnetic vertex in the diagrams of the off-diagonal thermoelectric
coefficient. An additional diagram appears due to modification of
the heat current vertex. According to \cite{RSWL} the heat current
vertex for noninteracting electrons is $\gamma_h=\xi_p  {\bf v}$,
which corresponds to the energy current measured with respect 
to the electron chemical potential. For electrons in the magnetic 
field the vertex of the heat current is modified according to Eq. (A.1)  
$$\gamma_h= \xi_p  {\bf v} + (e/c) ({\bf v} \cdot {\bf A}) {\bf v}. 
                                                                   \eqno(A.3)$$

Diagrams of LRM for the Nernst coefficient are presented in
Fig. 2. In the first and second diagrams the 
vertex $ (e/c) ({\bf v} \cdot {\bf A})$ describes the effect of the 
magnetic field on the electron states. In the third diagram the same
vertex is a part of the heat current operator, Eq. (A.3).
To get the term ${\bf A}({\bf k} \cdot \nabla T)$, one should expand
the Green function $G({\bf p}+ {\bf k}) $ in powers of 
$({\bf k}\cdot {\bf v})$.
Then the contribution of the first and second diagrams is given by
$$  N_1+N_2= {e\over TH}\int  
 {d^4 P \over (2\pi)^4} \ 
 { \partial S_0 (\epsilon) \over \partial \epsilon } \ \xi_p {\bf v}
{e \over c} ({\bf v} \cdot {\bf A}) ({\bf v} \cdot \nabla T) ({\bf v} 
\cdot {\bf k}) \ (I_1 + I_2),                                      \eqno(A.4)$$
where the combination of the Green functions is
$$ I_1+I_2 = 2i G^A(P) G^R(P)  {\rm Im} [G^A(P)]^2= 2i \tau^2  
{\rm Im} [G^A(P)]^2.                                             \eqno(A.5)$$ 

The contribution of the third diagram is 
$$ N_3= {e\over TH}\int  {d^4 P \over  (2\pi)^4} \  
{ \partial S_0 (\epsilon) \over \partial \epsilon } {e \over c} 
{\bf v}({\bf v} \cdot {\bf A})  ({\bf v} \cdot \nabla T) 
({\bf v} \cdot {\bf k})I_3 ,                                    \eqno(A.6)$$
where the combination of the Green functions $I_3$ is
$$ I_3 = 2i G^A(P) {\rm Im} [G^A(P)]^2=2i\tau^2 {\rm Im}G^A(P). \eqno(A.7)$$

The angular integral in Eqs. (A.4) and (A.6) gives 
the term proportional to  ${\bf A}({\bf k} \cdot \nabla T)$, 
$$ \int {d \Omega_{\bf p} \over 4\pi} \ {\bf v} ({\bf v} \cdot {\bf A})
 ({\bf v} \cdot {\bf k}) ({\bf v} \cdot \nabla T) = {v^4 \over 15}
{\bf A} ({\bf k} \cdot \nabla T ).                                 \eqno(A.8)$$
Therefore, the total contribution of three diagrams is
$$ N = {i e^2 \over 15 cTH} \int  {d \epsilon \over 2\pi} d\xi_p 
{ \partial S_0 (\epsilon) \over \partial \epsilon } \ v^4 \tau^2 \nu
\biggl(\xi_p {\rm Im} [G^A(P)]^2 +{\rm Im} G^A(P)\biggr) 
  {\bf A} ({\bf k} \cdot \nabla T ). 
                                                                 \eqno(A.9)$$

Without taking into account the particle-hole asymmetry the total
contribution of diagrams 1-3 goes to zero after integration
over $\xi_p$,
$$ \int d \xi_p \biggl( \xi_p {\rm Im}[G^A(P)]^2 +{\rm Im}[G^A(P)]
\biggr) =0.  \eqno(A.10)$$
Nonzero contribution arises from terms proportional to $\epsilon^2$, 
thus we should 
expand all electron parameters near the Fermi surface:
$$ [v(\xi_p)]^4 \nu (\xi_p) 
= v_0^4 \nu_0 \bigg[ 1+ {5\over 2}{\xi_p \over \epsilon_F}
+{15 \over 8}\bigg({\xi_p\over \epsilon_F} \bigg)^2+ ... \bigg],\eqno(A.11)$$ 
   
$$ [\tau(\epsilon)]^2=  
\tau_0^2 \bigg[1 -{\epsilon \over \epsilon_F}+...\bigg]         \eqno(A.12)$$
Taking into account terms proportional to
the square of the particle-hole asymmetry, e.g. $\xi^2/ \epsilon_F^2$ or
$ \xi \epsilon /\epsilon_F^2$, we get
$$ \int d \xi_p \ v^4 \nu \tau^2 \biggl(\xi_p {\rm Im}[G^A(P)^2 
+{\rm Im}G^A(P)\biggr)
= - \pi v_0^4 \nu_0 \tau_0^2 {5\over 4} {\epsilon^2 \over \epsilon_F^2}
= -\pi {5\over 2} {v_0^2 \tau_0^2 \nu_0 \over m} 
{\epsilon^2 \over \epsilon_F}.                                  \eqno(A.13)$$
Substituting this result into Eq. (A.9), and performing integration over
$\epsilon$, we get the Nernst coefficient which coincides with Eq. (48).

\begin{figure}
\caption{Diagrams of the kinetic electron self-energy $\Sigma^C$
contributing to the WL corrections to conductivity.}

\caption{Diagrams of the linear response method for the Nernst coefficient
of nonineracting electrons.}

\end{figure}
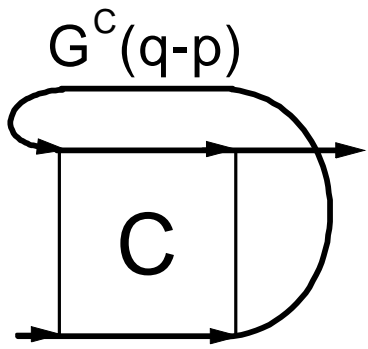
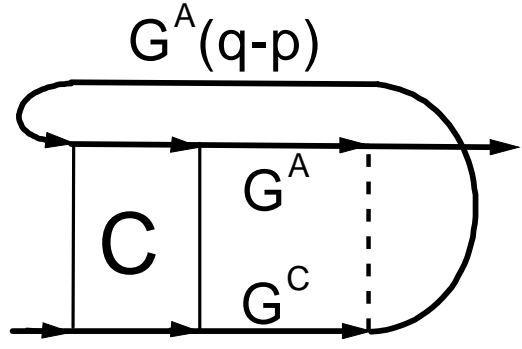
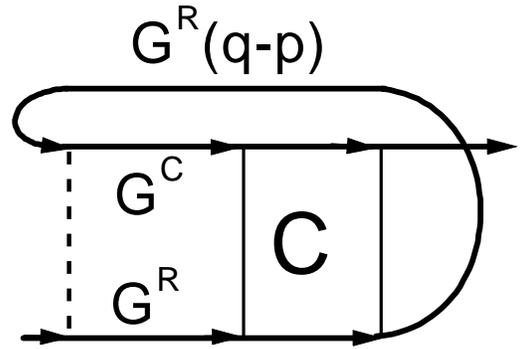
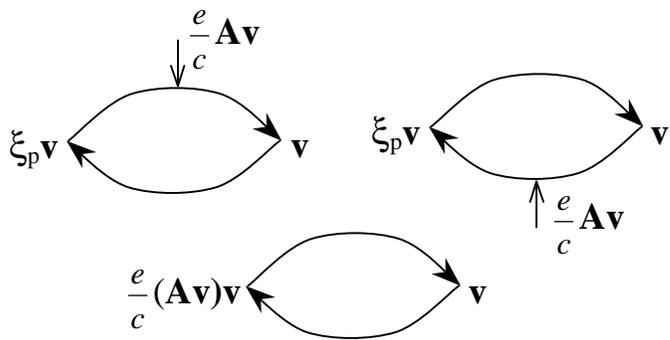
\end{document}